# Asymptotic Optimality of Antidictionary Codes

Takahiro Ota
Dept. of Electronic Engineering
Nagano Prefectural Institute of Technology
Ueda Nagano 386-1211, Japan
Email: ota@pit-nagano.ac.jp

Hiroyoshi Morita
Graduate School of Information Systems
University of Electro-Communications
Chofu Tokyo 182-8585, Japan
Email: morita@is.uec.ac.jp

*Abstract*—An antidictionary code is a lossless compression algorithm using an antidictionary which is a set of minimal words that do not occur as substrings in an input string. The code was proposed by Crochemore *et al.* in 2000, and its asymptotic optimality has been proved with respect to only a specific information source, called balanced binary source that is a binary Markov source in which a state transition occurs with probability 1/2 or 1. In this paper, we prove the optimality of both static and dynamic antidictionary codes with respect to a stationary ergodic Markov source on finite alphabet such that a state transition occurs with probability $p$ $(0 < p \leq 1)$.

## I. INTRODUCTION

This paper proves two theorems with respect to asymptotic optimality of both static and dynamic antidictionary codes for stationary ergodic Markov information sources. An antidictionary for a given string is a set of words of minimal length that never appear in the string, and it is in particular useful for data compression. An antidictionary coding scheme, called Data Compression using Antidictionaries (DCA), was first proposed by Crochemore *et al.* [1] for binary strings. Some extensions of the DCA, which are able to handle a finite alphabet and applied to arithmetic codes, have been proposed [2]–[4] (cf. [5]). Those algorithms work in an off-line manner, while some on-line DCA algorithms using dynamic suffix trees work with linear time and space have been proposed [6]–[8]. Moreover, a memory-efficient DCA using suffix arrays was proposed [9]. It was shown that the algorithm [8] achieves compression ratios as well as an efficient off-line data compression algorithm using Burrows-Wheeler transformation [10] by simulation results.

On the other hand, for only balanced binary sources, asymptotic optimality of a static DCA algorithm has been proved [1]. It was shown that the algorithm is asymptotically optimal for the source generated by an antidictionary if and only if the antidictionary is given to the algorithm in advance [1]. The averaged code length per symbol converges to the entropy rate of the source with probability one. The balanced binary source is a Markov source of finite order and emits all the strings which do not contain any word of the antidictionary as the substrings. Moreover, for any state of the Markov source with only one outgoing edge, probability one is assigned to each edge, while for that with two outgoing edges, probability 1/2 is assigned to those edges.

In this paper, we prove asymptotic optimality of a static and a dynamic DCA for a Markov source constructed from an antidictionary on finite alphabet such that a state transition occurs with probability $p$ $(0 < p \leq 1)$. This paper is organized as follows. Section II gives the basic definitions and notations. Section III shows review of the DCA algorithms. Section IV proves two theorems with respect to the asymptotic optimality of a static and a dynamic antidictionary code, respectively. Section V summarizes our results.

## II. BASIC DEFINITIONS AND NOTATIONS

Let $\mathcal{X} = \{0, 1, \ldots, J-1\}$ be a finite alphabet and $\mathcal{X}^*$ be the set of all finite strings over $\mathcal{X}$, including the null string of length zero, denoted by $\lambda$. For $\mathcal{X}$ and $\boldsymbol{x} \in \mathcal{X}^*$, $|\mathcal{X}|$ and $|\boldsymbol{x}|$ represent the size of $\mathcal{X}$ and the length of $\boldsymbol{x}$, respectively. For a string $\boldsymbol{x} = x_1 x_2 \ldots x_n \in \mathcal{X}^n$ of length $n$, let $\Sigma(\boldsymbol{x})$ be the set of all *suffixes* of $\boldsymbol{x}$, that is, $\Sigma(\boldsymbol{x}) = \{x_i x_{i+1} \ldots x_n | 1 \leq i \leq n\} \cup \{\lambda\}$, and let $\mathcal{D}(\boldsymbol{x})$ be the *dictionary* of all substrings of $\boldsymbol{x}$, that is, $\mathcal{D}(\boldsymbol{x}) = \{x_i x_{i+1} \ldots x_j | 1 \leq i \leq j \leq n\} \cup \{\lambda\}$. Let $\boldsymbol{x}^i$ be the *prefix* of length $i$ of $\boldsymbol{x}$, and we define that $\boldsymbol{x}^0 = \lambda$.

### A. Markov Source

Let $\mathcal{A} \subset \mathcal{X}^* \setminus \{\lambda\}$ be a non-empty finite set, and we assume that no word $\boldsymbol{u} \in \mathcal{A}$ is a substring of any $\boldsymbol{v} \in \mathcal{A}$ such as $\boldsymbol{v} \neq \boldsymbol{u}$. Crochemore *et al.* showed a deterministic automaton $F(\mathcal{A})$ which accepts all strings that contain no strings of $\mathcal{A}$ as their substrings [11]. In [1], $F(\mathcal{A})$ is used as an encoder and a decoder of static DCA algorithm. The set $\mathcal{A}$ will be referred to as the *antidictionary* and a string in $\mathcal{A}$ will be referred to as the *Minimal Forbidden Word (MFW)*. A deterministic automaton $F(\mathcal{A}) = (\mathcal{U}, \mathcal{X}, s_1, \mathcal{A})$ is defined as follows: Let $s(\boldsymbol{w})$ be the state corresponding to string $\boldsymbol{w}$ in $F(\mathcal{A})$. In other words, $s(\boldsymbol{w})$ is the state reached by string $\boldsymbol{w}$ from the initial state $s_1$.

- The initial state $s_1$ is $s(\lambda)$.
- A state $s(\boldsymbol{v})$ for $\boldsymbol{v} \in \mathcal{A}$ is called *sink state*. Any sink state has $|\mathcal{X}|$ outgoing edges, all having distinct labels, and all the edges of the state terminate the state.
- $\mathcal{U} = \{\boldsymbol{u} | \boldsymbol{u} \text{ is a proper prefix of } \boldsymbol{v} \in \mathcal{A}\}$. Note that a proper prefix of $\boldsymbol{v} = v_1 v_2 \ldots v_i$ is any of strings $v_1 v_2 \ldots v_j$ for $1 \leq j < i$, or $\lambda$. A state $s(\boldsymbol{u})$ has $|\mathcal{X}|$ outgoing edges, all having distinct labels. These edges are defined in the following manner: for each $a \in \mathcal{X}$,
  (i) if $\boldsymbol{u}a \in \mathcal{U}$, then the edge labeled $a$ from $s(\boldsymbol{u})$ terminates at $s(\boldsymbol{u}a)$.
  (ii) if $\boldsymbol{u}a \notin \mathcal{U}$, then the edge labeled $a$ from $s(\boldsymbol{u})$ terminates at $s(\boldsymbol{w})$, where $\boldsymbol{w}$ is the longest suffix of $\boldsymbol{u}a$ such as $\boldsymbol{w} \in (\mathcal{U} \cup \mathcal{A})$.

Let $G(\mathcal{A})$ be the automaton obtained by deleting from $F(\mathcal{A})$ all sink states and all edges incoming sink states. Fig. 1 shows $G(\mathcal{A})$ and $F(\mathcal{A})$, where $\mathcal{A} = \{11, 000, 10101\}$ and $\mathcal{X} = \{0, 1\}$. In Fig. 1, the solid lines and circles represent $G(\mathcal{A})$, while $G(\mathcal{A})$ with the dotted lines and squares represents $F(\mathcal{A})$, where squares represent sink states. To avoid trivial cases, we suppose that any state of $G(\mathcal{A})$ has at least one outgoing edge. For a state $s$ of $G(\mathcal{A})$, let $\mathcal{E}(s)$ be the set of labeled symbols of all outgoing edges from $s$.

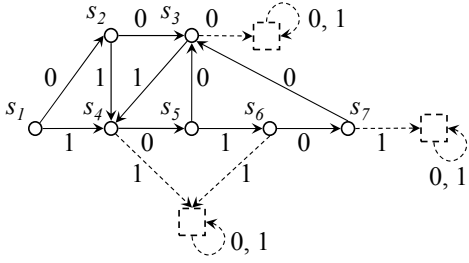

Fig. 1. The automaton $G(\mathcal{A})$ and $F(\mathcal{A})$ for $\mathcal{A} = \{11, 000, 10101\}$.

Let $S$ be the set of all states of $G(\mathcal{A})$, and let $S_1$ and $S_2$ be the set of all states having only one outgoing edge and that of all states having at least two outgoing edges, respectively. For $G(\mathcal{A})$, let $T : S \times \mathcal{X} \to S$ be transition probabilities independent of time called *transition probability matrix*. A stationary Markov (or unifilar, cf. [12]) source $\mathbf{X}_\mathcal{A}$ is characterized by $T$ of $G(\mathcal{A})$, and let $(\mu_1, \mu_2, \ldots, \mu_{|S|})$ be the stationary distribution whose components are the stationary probabilities of their states. We call $\mathbf{X}_\mathcal{A}$ *antidictionary source* in this paper.

Moreover, $\mathbf{X}_\mathcal{A}$ is a source called *shift of finite type* [13] since $\mathbf{X}_\mathcal{A}$ is described by a finite set of forbidden strings. Hence, $\mathbf{X}_\mathcal{A}$ is a stationary ergodic source [13]. A sequence $\mathbf{X}^n = X_1 X_2 \ldots X_n$ represents the sequence of random variables of length $n$ on $\mathbf{X}_\mathcal{A} = \{X_j : j = 1, 2, \ldots\}$. For a state $s_i$ of $G(\mathcal{A})$ in $\mathbf{X}_\mathcal{A}$, $p_{ic}$ represents the transition probability of the outgoing edge from $s_i$ with label $c$. The entropy $H(\mathbf{X}_\mathcal{A})$ is given by

$$H(\mathbf{X}_\mathcal{A}) = - \sum_{i:s_i \in S_2} \mu_i \sum_{c=0}^{|\mathcal{X}|-1} p_{ic} \log_2 p_{ic}, \quad (1)$$

where $0 \log_2 0 = 0$. Specially, if $\mathbf{X}_\mathcal{A}$ satisfies that $|\mathcal{X}| = 2$, $p_{j0} = p_{j1} = 1/2$ for any $s_j \in S_2$ and $p_{k0} = 1$ or $p_{k1} = 1$ for any $s_k \in S_1$, then $\mathbf{X}_\mathcal{A}$ is called *binary balanced source*.

The automaton $G(\mathcal{A})$ has a useful property, called *synchronization property* [1]. For a state $s_i$, let $l(s_i)$ be the locus string $\mathbf{u}$ such that $s_i = s(\mathbf{u})$ and $\mathbf{u} \in \mathcal{U}$ are satisfied. Notice that $s(l(s_i)) = s_i$.

Let $\mathbf{u}$ and $\mathbf{v}$ be the string $l(s_i)$ and $l(s_j)$ for states $s_i$ and $s_j$ ($i \neq j$), respectively, and let $m$ be length of the longest MFW in $\mathcal{A}$. Then, we have the following theorem.

*Theorem A (Theorem 3 [1]):* For any string $\mathbf{w} \in \mathcal{X}^*$ of length $m-1$, if both strings $\mathbf{uw}$ and $\mathbf{vw}$ do not contain any string of $\mathcal{A}$ as the substrings, then $s(\mathbf{uw}) = s(\mathbf{vw})$.

In other words, suppose that $s_d$ and $s_e$ are the states reached by $\mathbf{w}$ from $s_i$ and $s_j$, respectively, so that $s_d = s_e$ if the conditions are satisfied shown in Theorem A. In Fig. 1, $m-1$ is given by 4 since length of the longest MFW, that is, 10101, is 5. As an example, for $s_1$, $s_5$ and $\mathbf{w} = 0100$, the states reached by $\mathbf{w}$ from $s_1$ and $s_5$ are the same state $s_3$.

### B. Suffix Tree

The suffix tree of $\mathbf{x}$ is a tree structure [14] that stores all elements of $\Sigma(\mathbf{x})$. Let $\mathbb{T}_i$ be the suffix tree of $\mathbf{x}^i$. The string associated with the path from the root $\rho$ to a node $p$ in $\mathbb{T}_i$ is denoted by $\mathbf{w}(p)$, and we define that $\mathbf{w}(\rho)$ is $\lambda$. The string length $|\mathbf{w}(p)|$ will be referred to the *depth* of $p$. For any node $p$ in $\mathbb{T}_i$, let $\mathcal{L}_i(p)$ be the set of labeled symbols of all edges sprouting from $p$, that is, $\mathcal{L}_i(p) = \{a | \mathbf{w}(p)a \in \mathcal{D}(\mathbf{x}^i), a \in \mathcal{X}\}$. For any node $p \neq \rho$, we can write $\mathbf{w}(p) = a\mathbf{v}$, where $a \in \mathcal{X}$ and $\mathbf{v} \in \mathcal{X}^*$. Let $q$ be the node such that $\mathbf{w}(q) = \mathbf{v}$, and a pointer from $p$ to $q$, denoted by $\sigma(p)$, is called *suffix link*. For a given depth $d \geq 0$, if $|\mathbf{w}(p)| \geq d$, then let $\sigma_d(p)$ be a node of depth $d$ pointed by one of a series of suffix links starting from $p$ and moving back to the root $\rho$.

*Definition 1 (active point):* An *active point* $\alpha_i$ in $\mathbb{T}_i$ is the node corresponding to the string $\mathbf{u}$ such that the longest string in $(\Sigma(\mathbf{x}^i) \cap \mathcal{D}(\mathbf{x}^{i-1}))$ where $\alpha_0$ is the root $\rho$.

The active point plays a key roll in the on-line algorithm, called the Ukkonen algorithm, for constructing suffix trees with the linear complexity [15].

## III. REVIEW OF THE DCA ALGORITHMS

First, we describe a static DCA algorithm [4]. We suppose that Assumption 1 is satisfied for the static DCA algorithm.

*Assumption 1:* The static DCA algorithm knows $\mathcal{A}$.

From Assumption 1, notice that $G(\mathcal{A})$ plays as the encoder / decoder parts of the algorithm since $G(\mathcal{A})$ is constructed from $\mathcal{A}$. Table I shows output for $x_{i+1}$ in the static DCA algorithm. In Case-(1), no symbol is output, that is, $x_{i+1}$ is predictable since there exists only one outgoing edge from $s(\mathbf{x}^i)$. In Case-(2), $e(\cdot)$ represents an adaptive arithmetic coder of order-0 (cf. [16]). The probability $\Pr(x_{i+1}|s(\mathbf{x}^i))$ is calculated by $N(x_{i+1}|s(\mathbf{x}^i))/\sum_{c \in \mathcal{X}} N(c|s(\mathbf{x}^i))$, where $N(c|s(\mathbf{x}^i))$ is a counter that has the number of traversed times from $s(\mathbf{x}^i)$ with symbol $c$. Note that for $s_k$, if $c \in \mathcal{E}(s_k)$, then the initial value of $N(c|s_k)$ is set to 1. Otherwise its initial value is 0. For a given input string $\mathbf{x}$ of length $n$, the codeword of the static DCA algorithm is given by the triplet, that is,

$$(\mathcal{A}, e(\mathbf{x}), n). \quad (2)$$

TABLE I
OUTPUT FOR $x_{i+1}$ IN THE STATIC DCA ALGORITHM.

| Case | $|\mathcal{E}(s(\mathbf{x}^i))|$ | Output |
|---|---|---|
| (1) | 1 | none |
| (2) | at least 2 | $e(\Pr(x_{i+1}|s(\mathbf{x}^i)))$ |

Next, we describe a dynamic DCA algorithm [8]. The algorithm uses a subtree of the dynamic suffix tree, which has a given fixed depth $d+1$ ($d \geq 0$). In [8], a node $\beta_i$ in $\mathbb{T}_i$, called *modified active point*, is used to encode symbol $x_{i+1}$. The node $\beta_i$ is defined as follows:

*Definition 2 (modified active point):* For a given fixed integer $d \geq 0$,

$$\beta_i = \begin{cases} \alpha_i & (|\mathbf{w}(\alpha_i)| < d), \\ \sigma_d(\alpha_i) & (|\mathbf{w}(\alpha_i)| \geq d). \end{cases} \quad (3)$$

Table II shows the output for $x_{i+1}$ in the dynamic DCA algorithm. In Case-(0), the pair $(I, \mathsf{R}(x_{i+1}))$ is output, where $I$ represents an interval of insertion of new edge, and $\mathsf{R}(x_{i+1})$ represents the rank of $x_{i+1}$ ($1 \leq \mathsf{R}(x_{i+1}) \leq |\mathcal{X}|$). Let $\mathcal{L}_i(\beta_i)$ be a set $\{a | \mathbf{w}(\beta_i)a \in \mathcal{D}(\mathbf{x}^i), a \in \mathcal{X}\}$. Let $\mathcal{R}_i$ be a set of the longest string $\mathbf{w}(p)c$ in $(\Sigma(\mathbf{w}(\beta_i)c) \cap \mathcal{D}(\mathbf{x}^i))$ or $\{c\}$ for each $c \in (\mathcal{X} \setminus \mathcal{L}_i(\beta_i))$. Suppose that $\mathbf{w}(p)a, \mathbf{w}(q)b \in \mathcal{R}_i, a \neq b$.

TABLE II
OUTPUT FOR $x_{i+1}$ IN THE DYNAMIC DCA ALGORITHM.

| Case | Relationship between $\beta_i$ and $x_{i+1}$ | Output |
|---|---|---|
| (0) | $x_{i+1} \notin \mathcal{L}_i(\beta_i)$ | $(I, \mathsf{R}(x_{i+1}))$ |
| (1) | $|\mathcal{L}_i(\beta_i)| = 1$ and $x_{i+1} \in \mathcal{L}_i(\beta_i)$ | none |
| (2) | $|\mathcal{L}_i(\beta_i)| \geq 2$ and $x_{i+1} \in \mathcal{L}_i(\beta_i)$ | $e(\Pr(x_{i+1}|\beta_i))$ |

If a following condition in (4), (5) and (6) is satisfied, then $\mathsf{R}(a) < \mathsf{R}(b)$.

$$|\boldsymbol{w}(p)a| > |\boldsymbol{w}(q)b|, \tag{4}$$
$$|\boldsymbol{w}(p)a| = |\boldsymbol{w}(q)b| \text{ and } N(a|p) > N(b|q), \tag{5}$$
$$|\boldsymbol{w}(p)a| = |\boldsymbol{w}(q)b|, \ N(a|p) = N(b|q) \text{ and}$$
$$a < b \text{ (in lexicographical)}, \tag{6}$$

where $N(\cdot|\cdot)$ is a counter used in Case-(2). The rank $\mathsf{R}(x_{i+1})$ is determined by traversing up suffix links starting from $\beta_i$ to $\rho$ and is the rank of the string which has $x_{i+1}$ as the last symbol in $\mathcal{R}_i$. The rank $\mathsf{R}(x_{i+1})$ is used to convert $x_{i+1}$ into a small integer to improve the compression ratio. The reason is that a symbol $c \in (\mathcal{X} \backslash \mathcal{L}_i(\beta_i))$ having high probability will be found at a node near $\beta_i$ on the suffix links. The details are described in [7].

In Case-(1), no symbol is output since $x_{i+1}$ is predictable from the fact that there exists only one edge from $\beta_i$. In Case-(2), the probability $\Pr(x_{i+1}|\beta_i)$ is calculated by $N(x_{i+1}|\beta_i)/\sum_{c \in \mathcal{X}} N(c|\beta_i)$, where $N(c|\beta_i)$ is a counter that has the number of traversed times from the internal node $\beta_j$ with symbol $c$ ($0 \leq j \leq i-1$). Note that for an internal node $n_k$ of $\mathbb{T}_i$ such as $|\mathcal{L}_i(n_k)| \geq 2$, if $c \in \mathcal{L}_i(n_k)$, then the initial value of $N(c|n_k)$ is set to 1. Otherwise its initial value is 0.

Let $l_n^s$ be the codeword length per symbol of the static DCA algorithm for a random string of length $n$. That is, $l_n^s$ is given by (the codeword length)/$n$. Then, the following theorem holds.

*Theorem B:* [Theorem 7 [1]] Under Assumption 1, for a balanced binary source $\mathbf{X}_\mathcal{A}$, $l_n^s$ converges to $H(\mathbf{X}_\mathcal{A})$ with probability one as $n \to \infty$.

## IV. MAIN RESULTS

If $\mathbf{X}_\mathcal{A}$ is stationary ergodic, then we obtain the following theorem for the static DCA algorithm.

*Theorem 1:* Under Assumption 1, for a stationary ergodic source $\mathbf{X}_\mathcal{A}$, $l_n^s$ converges to $H(\mathbf{X}_\mathcal{A})$ with probability one as $n \to \infty$.

Now, let $l_n^d$ be the codeword length per symbol of the dynamic DCA algorithm for a random string of length $n$. And let $m$ be the length of the longest MFW in $\mathcal{A}$. Moreover, we have the following assumption on the dynamic DCA algorithm.

*Assumption 2:* Both encoder and decoder of the dynamic DCA algorithm do not know $\mathcal{A}$ while they know $m$.

*Theorem 2:* Under Assumption 2, for a stationary ergodic source $\mathbf{X}_\mathcal{A}$, $l_n^d$ converges to $H(\mathbf{X}_\mathcal{A})$ with probability one as $n \to \infty$.

### A. Proof of Theorem 1

We use three lemmas to prove Theorem 1. Let $S_{2,0}$ and $S_{2,\infty}$ be the set of states in $S_2$ for $\mu_i = 0$ and $\mu_i > 0$, respectively. For $\mathbf{X}^n$, let $Y_{i,n}$ be a random variable taking values in the number of traversed times of $s_i$, and let $|e(\mathbf{X}^n)|$ be a random variable taking values in the length of output of Case-(2), that is $e(\boldsymbol{x})$ in (2). For a given symbol $c \in \mathcal{X}$ and $s_i \in S_{2,\infty}$, let $Z_{ic,h}$ be a random variable, when $s_i$ is traversed at the $h$th time, such as

$$Z_{ic,h} = \begin{cases} 1 & (z = c), \\ 0 & (z \neq c), \end{cases} \tag{7}$$

where $z$ is the labeled symbol of traversed outgoing edge from $s_i$ at the time. For a positive integer $k$, $[Z_{ic}]_k$ is given by $[Z_{ic}]_k = (Z_{ic,1} + Z_{ic,2} + \cdots + Z_{ic,k})/k$.

*Lemma 1:* $\Pr\{\lim_{n \to \infty} Y_{i,n}/n = \mu_i\} = 1$.
*Lemma 2:* $\Pr\{\lim_{n \to \infty} [Z_{ic}]_{(Y_{i,n})} = p_{ic}\} = 1$.
*Lemma 3:* $\Pr\{\limsup_{n \to \infty} |e(\mathbf{X}^n)|/n = H(\mathbf{X}_\mathcal{A})\} = 1$.

*(Proof of Lemma 1):* From the definition of $\mathbf{X}_\mathcal{A}$, the steady state probability of $s_i$ is given by $\mu_i$. Therefore, the lemma holds. ∎

*(Proof of Lemma 2):* A sequence $\mathbf{Z} = Z_{ic,1} Z_{ic,2} \ldots$ is i.i.d. and $Z_{ic,h}$ ($h = 1, 2, \ldots$) has the same probability distribution. And, from the definition of $Z_{ic,h}$, the expected value $\mathbb{E}(Z_{ic,h})$ equals to $p_{ic}$. Moreover, for $s_i \in S_{2,\infty}$, from Lemma 1, $Y_{i,n}$ diverges to infinity as $n \to \infty$ with prob. 1. Therefore, from the strong law of large numbers, the lemma holds. ∎

*(Proof of Lemma 3):*

$$\limsup_{n \to \infty} \frac{|e(\mathbf{X}^n)|}{n}$$
$$= -\limsup_{n \to \infty} \frac{1}{n} \sum_{i: s_i \in S_2} Y_{i,n} \sum_{c=0}^{|\mathcal{X}|-1} [Z_{ic}]_{(Y_{i,n})} \log_2 [Z_{ic}]_{(Y_{i,n})} \tag{8}$$
$$\stackrel{(a)}{=} -\sum_{i: s_i \in S_2} \limsup_{n \to \infty} \frac{Y_{i,n}}{n} \sum_{c=0}^{|\mathcal{X}|-1} [Z_{ic}]_{(Y_{i,n})} \log_2 [Z_{ic}]_{(Y_{i,n})} \tag{9}$$
$$= -\sum_{i: s_i \in S_{2,0}} \limsup_{n \to \infty} \frac{Y_{i,n}}{n} \sum_{c=0}^{|\mathcal{X}|-1} [Z_{ic}]_{(Y_{i,n})} \log_2 [Z_{ic}]_{(Y_{i,n})}$$
$$- \sum_{i: s_i \in S_{2,\infty}} \limsup_{n \to \infty} \frac{Y_{i,n}}{n} \sum_{c=0}^{|\mathcal{X}|-1} [Z_{ic}]_{(Y_{i,n})} \log_2 [Z_{ic}]_{(Y_{i,n})} \tag{10}$$
$$\stackrel{(b)}{=} -\sum_{i: s_i \in S_{2,\infty}} \mu_i \sum_{c=0}^{|\mathcal{X}|-1} p_{ic} \log_2 p_{ic} \tag{11}$$
$$\stackrel{(c)}{=} H(\mathbf{X}_\mathcal{A}), \tag{12}$$

where $(a)$ follows from the fact that an index $i$ of state in $G(\mathcal{A})$ is independent of $n$, and $(b)$ follows that addition to Lemma 1, Lemma 2 and the first term of right-hand side of (10) converges to 0 with prob. 1 as $n \to \infty$ since $\mu_i = 0$ for any $s_i \in S_{2,0}$, and $(c)$ follows from (1). ∎

*(Proof of Theorem 1):* From (2), $l_n^s$ is given by

$$l_n^s \leq \limsup_{n \to \infty} \left( \frac{\#\mathcal{A}}{n} + \frac{|e(\mathbf{X}^n)|}{n} + \frac{|\omega^*(n)|}{n} \right), \tag{13}$$

where $\#\mathcal{A}$ is a size of list of all the MFWs in $\mathcal{A}$, and $\omega^*(n)$ is a representation of $n$ using the Elias $\omega^*$ code for positive integers [17] (cf. [12]). The length $|\omega^*(n)|$ is given by

$$|\omega^*(n)| \leq \log_2 n + 2\log_2(\log_2 n) + 7. \tag{14}$$

From (14), the third term of the right-hand side of (13) converges to 0 as $n \to \infty$. From Assumption 1, $\#\mathcal{A}$ is a constant, so that the first term also converges to 0 as $n \to \infty$. Therefore, from Lemma 3, the theorem holds with prob. 1. ∎

## B. Proof of Theorem 2

We use eight lemmas to prove Theorem 1. For a given fixed integer $m \geq 1$ in Assumption 2, we use $m-1$ as the depth $d$ in Definition 2, that is

$$d = m - 1. \tag{15}$$

We define a random variable $V_k \stackrel{\text{def}}{=} X_k X_{k+1} \ldots X_{d+k} \in \mathcal{X}^{d+1}$ for $k \geq 1$. For $V_k$, a random variable $Q_k$ is defined as

$$Q_k = \begin{cases} 0 & (\exists i : V_k = v = V_i, (1 \leq i \leq k-1)), \\ 1 & (V_k = v \neq V_i, (1 \leq \forall i \leq k-1), \end{cases} \tag{16}$$

where $v$ is a string satisfying that $\Pr\{V_1 = v\} > 0$. Note that we define that $Q_1$ takes value 1. For a string $\boldsymbol{x}^n$ on $\mathbf{X}_{\mathcal{A}}$, let $\Delta_n$ be the set of all nodes whose depth is $d$ in $\mathbb{T}_n$, and for any state $s_j$ $(1 \leq j \leq |S|)$ of $G(\mathcal{A})$, we define that $\Delta_{j,n} = \{p \mid s_j = s(\boldsymbol{w}(p)), p \in \Delta_n\}$. Note that for a node $p \in \Delta_n$, the unique state of $G(\mathcal{A})$ is determined from Theorem A since $|\boldsymbol{w}(p)| = d$ and $d = m - 1$.

For a node $p$, let $N_n(p)$ be the random number of times $\beta_h$ passed $p$ $(0 \leq h \leq n-1)$. For a given symbol $c \in \mathcal{X}$ and $p \in \Delta_{j,n}$, let $\tilde{Z}_{jc,k}$ be a random variable, when $p$ is traversed at the $k$th time, such as

$$\tilde{Z}_{jc,k} = \begin{cases} 1 & (z = c), \\ 0 & (z \neq c), \end{cases} \tag{17}$$

where $z$ is the labeled symbol of traversed edge from $p$ at the time. For a positive integer $g$, $[\tilde{Z}_{jc}]_g$ is given by $[\tilde{Z}_{jc}]_g = (\tilde{Z}_{jc,1} + \tilde{Z}_{jc,2} + \cdots + \tilde{Z}_{jc,g})/g$. Let $D_n$ be a random variable taking the depth of $\beta_n$, that is $|\boldsymbol{w}(\beta_n)|$, and let $E_n$ be a random variable taking the index of $s(\boldsymbol{w}(\beta_n))$.

*Lemma 4:* If $x_{n-d+1} x_{n-d+2} \ldots x_n \in \mathcal{D}(\boldsymbol{x}^{n-1})$, then $|\boldsymbol{w}(\beta_n)| = d$.

*Lemma 5:* If $\beta_n \in \Delta_n$, then $s(\boldsymbol{w}(\beta_n)) = s(\boldsymbol{x}^n)$.

*Lemma 6:* $\Pr\{\lim_{n \to \infty} Q_n = 0\} = 1$.

*Lemma 7:* $\Pr\{\lim_{n \to \infty} D_n = d\} = 1$.

*Lemma 8:* $\Pr\{\lim_{n \to \infty} E_n = s(\boldsymbol{x}^n)\} = 1$.

*Lemma 9:* $\Pr\{\lim_{n \to \infty} \sum_{p \in \Delta_{j,n}} N_n(p)/n = \mu_j\} = 1$.

*Lemma 10:* For $p \in \Delta_{j,n}$, $\Pr\{\lim_{n \to \infty} \mathcal{L}_n(p) = \mathcal{E}(s_j)\} = 1$.

*Lemma 11:* For $p \in \Delta_{j,n}$, $\Pr\{\lim_{n \to \infty} [\tilde{Z}_{jc}]_{(N_n(p))} = p_{jc}\} = 1$.

*(Proof of Lemma 4):* Since $\boldsymbol{v} = x_{n-d+1} x_{n-d+2} \ldots x_n \in \Sigma(\boldsymbol{x}^n)$, we have $\boldsymbol{v} \in (\Sigma(\boldsymbol{x}^n) \cap \mathcal{D}(\boldsymbol{x}^{n-1}))$. From Definition 1, we obtain $|\boldsymbol{w}(\alpha_n)| \geq |\boldsymbol{v}| = d$. Therefore, we have $|\boldsymbol{w}(\beta_n)| = d$ from (3). ∎

*(Proof of Lemma 5):* Since $\beta_n \in \Delta_n$, we have $\boldsymbol{w}(\beta_n) = \boldsymbol{w} = x_{n-d+1} x_{n-d+2} \ldots x_n$ and $|\boldsymbol{x}^n| \geq |\boldsymbol{w}| = d$. From Theorem A and $s(\boldsymbol{w}) = s(\boldsymbol{x}^{n-d} \boldsymbol{w})$, we have $s(\boldsymbol{w}(\beta_n)) = s(\boldsymbol{x}^n)$. ∎

*(Proof of Lemma 6):* Since $\mathbf{X}_{\mathcal{A}}$ is a stationary ergodic source, the lemma holds (cf. [18]). ∎

*(Proof of Lemma 7):* Since $d$ is a constant, $\Pr\{\lim_{n \to \infty} Q_{n-d+1} = 0\} = 1$ from Lemma 6. Therefore, there exists $j$ $(1 \leq j \leq n-d+1)$ such that $X_{n-d+1} X_{n-d+2} \ldots X_n = X_j X_{j+1} \ldots X_{j+d-1}$ with probability 1. Hence from Lemma 4, the lemma holds. ∎

*(Proof of Lemma 8):* From Lemmas 5 and 7, the lemma holds. ∎

*(Proof of Lemma 9):* Suppose that $s(\boldsymbol{x}^n) = s_j$. From Lemma 5, if $\beta_n \in \Delta_n$, then $\beta_n \in \Delta_{j,n}$, that is $s(\boldsymbol{w}(\beta_n)) = s_j$. On the other hand, if $\beta_n \notin \Delta_n$, that is $|\boldsymbol{w}(\beta_n)| < d$, then $s(\boldsymbol{w}(\beta_n)) \neq s_j$ can hold.

We evaluate that the maximum total number $M$ of occurrences such that $|\boldsymbol{w}(\beta_k)| < d$ for $0 \leq k \leq n-1$. Let $\boldsymbol{w}$ be the suffix of $\boldsymbol{x}^n$ of length $d$. If $\boldsymbol{w} \notin \mathcal{D}(\boldsymbol{x}^{n-1})$, then $|\boldsymbol{w}(\beta_n)| < d$ from Definition 2. On the other hand, if all the strings, whose lengths are not more than $d$, are included in $\mathcal{D}(\boldsymbol{x}^{n-1})$, then $|\boldsymbol{w}(\beta_n)| = d$. Therefore, $M$ is the total number of strings in $\mathcal{X}^*$, whose length are not more than $d$ since $\mathcal{D}(\boldsymbol{x}^{n-1})$ is monotone increasing with respect to $n$. Hence, $M$ is given by $(|\mathcal{X}|^{d+1} - 1)/(|\mathcal{X}| - 1)$ for $|\mathcal{X}| \geq 2$. In other words, it is equal to the number of nodes of a tree, called $|\mathcal{X}|$-ary tree, such that any external node has depth $d$ and any internal node has exactly $|\mathcal{X}|$ descendants (cf. [19]). Note that for $|\mathcal{X}| = 1$, the total number is given by $d$. By using $M$, for any $\Delta_{j,n}$, the following equation holds.

$$\frac{Y_{j,n}}{n} - \frac{M}{n} \leq \sum_{p \in \Delta_{j,n}} \frac{N_n(p)}{n} \leq \frac{Y_{j,n}}{n} + \frac{M}{n}. \tag{18}$$

Since $|\mathcal{X}|$ and $d$ are constants, $M$ is a constant. Hence, the term $M/n$ converges to 0 as $n \to \infty$. Therefore, $Y_{j,n}/n$ converges to $\mu_j$ as $n \to \infty$ from Lemma 1, so that the lemma holds. ∎

*(Proof of Lemma 10):* Due to $p \in \Delta_{j,n}$, we have $s(\boldsymbol{w}(p)) = s_j$. Hence, $\limsup_{n \to \infty} \mathcal{L}_n(p) = \mathcal{E}(s_j)$. Next, we will show that $\Pr\{\liminf_{n \to \infty} \mathcal{L}_n(p) = \mathcal{E}(s_j)\} = 1$. For a string $\boldsymbol{x}^n$, $\mathcal{D}(\boldsymbol{x}^n)$ is monotone increasing with respect to $n$, so that we have $\mathcal{L}_n(p) \subseteq \mathcal{L}_{n+1}(p)$. Moreover, from Lemma 6, for any $c \in \mathcal{E}(s_j)$, $\boldsymbol{w}(p)c \in \mathcal{D}(\boldsymbol{x}^{n-1})$ as $n \to \infty$ with prob. 1. Therefore, $\Pr\{\liminf_{n \to \infty} \mathcal{L}_n(p) = \mathcal{E}(s_j)\} = 1$. Hence, the lemma holds. ∎

*(Proof of Lemma 11):* Due to $p \in \Delta_{j,n}$, we have $s(\boldsymbol{w}(p)) = s_j$. Therefore, from Lemmas 6 and 10, $\tilde{Z}_{jc,k}$ has the same probability distribution of $Z_{jc,k}$, and $\mathbb{E}(\tilde{Z}_{jc,k})$ equals to $\mathbb{E}(Z_{jc,k})$ $(k = 1, 2, \ldots)$. Hence, $\mathbb{E}(\tilde{Z}_{jc,k})$ equals to $p_{jc}$. Moreover, a sequence $\tilde{\mathbf{Z}} = \tilde{Z}_{jc,1} \tilde{Z}_{jc,2} \ldots$ is i.i.d. Since $\mathbf{X}_{\mathcal{A}}$ is supposed to be a stationary ergodic source and $\Pr\{V_1 = \boldsymbol{w}(p)\} > 0$, $N_n(p)$ diverges to infinity with prob. 1 as $n \to \infty$. Therefore, from the strong law of large numbers, the lemma holds. ∎

*(Proof of Theorem 2):* Let $C(\mathbf{X}^n)$ be the codeword length achieved by the dynamic DCA algorithm, and let $C_0(\mathbf{X}^n)$ and $C_2(\mathbf{X}^n)$ be the codeword length in Case-(0) and Case-(2), respectively, that is,

$$C(\mathbf{X}^n) = C_0(\mathbf{X}^n) + C_2(\mathbf{X}^n). \tag{19}$$

Therefore, $l_n^d$ is given by

$$l_n^d = \lim_{n \to \infty} \frac{C(\mathbf{X}^n)}{n}. \tag{20}$$

First, we evaluate $C_0(\mathbf{X}^n)$. Let $n_0$ be the total number of occurrences of Case-(0) for a given $\boldsymbol{x}^n$ on $\mathbf{X}_{\mathcal{A}}$, and let $I_0$ and $R_0$ be the maximum code length of $I$ and $\mathsf{R}(x_{i+1})$ shown in Table II for $0 \leq i \leq n-1$. Suppose that $|\mathcal{X}| \geq 2$. The value $n_0$ is not more than the total number of strings whose length is not more than $d+1$ in $\mathcal{X}^*$, since Case-(0) occurs if $\boldsymbol{w}(\beta_i) x_{i+1} \notin \mathcal{D}(\boldsymbol{x}^i)$. Therefore, we obtain

$$n_0 \leq (|\mathcal{X}|^{d+2} - 1)/(|\mathcal{X}| - 1). \tag{21}$$

Moreover, the maximum length of $I$ is $n$. Hence, by using Elias $\omega^*$ code, we obtain

$$I_0 \leq |\omega^*(n)|. \tag{22}$$

By using a fixed length code for a symbol with respect to $\mathsf{R}(x_{i+1})$,

$$R_0 = \log_2 |\mathcal{X}|. \tag{23}$$

From (21), (22), and (23),

$$C_0(\boldsymbol{x}^n)/n \leq n_0 \cdot (I_0 + R_0)/n \tag{24}$$
$$\leq \frac{(|\mathcal{X}|^{d+2} - 1) \cdot (|\omega^*(n)| + \log_2 |\mathcal{X}|)}{(|\mathcal{X}| - 1) \cdot n}. \tag{25}$$

Since $|\mathcal{X}|$ and $d$ are constants, from (14), $C_0(\boldsymbol{x}^n)/n$ converges to 0 as $n \to \infty$. Therefore,

$$\lim_{n \to \infty} C_0(\boldsymbol{x}^n)/n = 0. \tag{26}$$

Note that in case of $|\mathcal{X}| = 1$, since $n_0 \leq d+1$ and $I_0 = R_0 = 1$, equation (26) holds.

Next, we evaluate $C_2(\mathbf{X}^n)$. For a given $\boldsymbol{x}^n$, let $l(p)$ be the averaged code length of Case-(2) for a node $p$ in $\mathbb{T}_n$. Note that $l(p) < \infty$ since $|\mathcal{X}|$ is finite. We have

$$\lim_{n \to \infty} \frac{C_2(\boldsymbol{x}^n)}{n} \leq \limsup_{n \to \infty} \frac{1}{n} \sum_{|\mathcal{L}_n(p)| \geq 2} N_n(p) l(p) \tag{27}$$

$$= \limsup_{n \to \infty} \frac{1}{n} \sum_{|\mathcal{L}_n(p)| \geq 2, p \notin \Delta_n} N_n(p) l(p)$$

$$+ \limsup_{n \to \infty} \frac{1}{n} \sum_{|\mathcal{L}_n(p)| \geq 2, p \in \Delta_n} N_n(p) l(p). \tag{28}$$

For $p \notin \Delta_n$, the maximum value of $N_n(p)$ is less than or equal to the total number $M$ of strings whose lengths are not more than $d$ in $\mathcal{X}^*$, that is $M$ described in the proof of Lemma 9. Therefore, the first term in the right-hand side of (28) converges to 0 as $n \to \infty$ since $M$ is a constant. Let $\varepsilon_n$ be the first term. From (28), we obtain

$$\lim_{n \to \infty} \frac{C_2(\boldsymbol{x}^n)}{n} \leq \varepsilon_n + \limsup_{n \to \infty} \frac{1}{n} \sum_{j: s_j \in S_2} \sum_{p \in \Delta_{j,n}} N_n(p) l(p)$$

$$\stackrel{(a)}{=} \varepsilon_n + \sum_{j: s_j \in S_{2,0}} \limsup_{n \to \infty} \frac{1}{n} \sum_{p \in \Delta_{j,n}} N_n(p) l(p)$$

$$+ \sum_{j: s_j \in S_{2,\infty}} \limsup_{n \to \infty} \frac{1}{n} \sum_{p \in \Delta_{j,n}} N_n(p) l(p), \tag{29}$$

where $(a)$ follows from the fact that an index $j$ of state of $G(\mathcal{A})$ is independent of $n$. From Lemma 9, the first term of right-hand side of (29) converges to 0 since $\mu_j = 0$ for $s_j \in S_{2,0}$ as $n \to \infty$. Let $\varepsilon'_n$ be the first term and let $\varepsilon''_n$ be $\varepsilon''_n = \varepsilon_n + \varepsilon'_n$. From (29),

$$\lim_{n \to \infty} \frac{C_2(\boldsymbol{x}^n)}{n} \leq \varepsilon''_n + \sum_{j: s_j \in S_{2,\infty}} \limsup_{n \to \infty} \frac{1}{n} \sum_{p \in \Delta_{j,n}} N_n(p) l(p). \tag{30}$$

For $p \in \Delta_{j,n}$, $l(p)$ is written by

$$l(p) = -\sum_{c=0}^{|\mathcal{X}|-1} [\tilde{Z}_{jc}]_{(N_n(p))} \log_2 [\tilde{Z}_{jc}]_{(N_n(p))}. \tag{31}$$

Moreover, for $p \in \Delta_{j,n}$, from Lemma 11 and (31),

$$l(p) = -\sum_{c=0}^{|\mathcal{X}|-1} p_{jc} \log_2 p_{jc} \tag{32}$$

with prob. 1 as $n \to \infty$. From (30), (32), and Lemma 9,

$$\lim_{n \to \infty} \frac{C_2(\boldsymbol{x}^n)}{n} \leq \varepsilon''_n - \sum_{j: s_j \in S_{2,\infty}} \mu_j \sum_{c=0}^{|\mathcal{X}|-1} p_{jc} \log_2 p_{jc} \tag{33}$$

with prob. 1. From (33) and (1),

$$\lim_{n \to \infty} \frac{C_2(\boldsymbol{x}^n)}{n} \leq \varepsilon''_n + H(\mathbf{X}_\mathcal{A}) \tag{34}$$

with prob. 1. From (19), (20), (26), (34), and since $\varepsilon''_n$ converges to 0 with prob. 1 as $n \to \infty$, we obtain

$$l_n^d = H(\mathbf{X}_\mathcal{A}) \tag{35}$$

with prob. 1 as $n \to \infty$. Therefore, the theorem holds. ∎

## V. CONCLUSION

In this paper, we proved asymptotic optimality of both static and dynamic DCA algorithms with respect to antidictionary sources, that is a stationary ergodic Markov source driven by $G(\mathcal{A})$. The averaged code length per symbol of the algorithms converge to the entropy rate of the source with probability one.